\documentclass[twoside,12pt]{article}
\usepackage{epsfig}
\usepackage[usenames]{color}
\usepackage{amssymb}
\usepackage[intlimits]{amsmath}
\usepackage{amsfonts}
\usepackage{subfigure}
\usepackage{graphicx}

\def\Journal#1#2#3#4{{#1} {#2} (#4) #3 }

\def\PRC{{\em Phys. Rev.} C}

\newcommand{\be}{\begin{equation}}
\newcommand{\ee}{\end{equation}}
\newcommand{\bea}{\begin{eqnarray}}
\newcommand{\eea}{\end{eqnarray}}

\newcommand{\sh}[1]{#1\hskip-7pt \diagup}
\newcommand{\Sh}[1]{#1\hskip-10pt \diagup}

\begin{document}

\title{Hadronic contribution to the muon $g-2$: a Dyson-Schwinger perspective}
\author{T.\ Goecke,$^{1}$ C.\ S.\ Fischer$^{1,2}$ R.\ Williams,$^3$ \\
\\
$^1$ Institut f\"ur Theoretische Physik, 
 Universit\"at Giessen, 35392 Giessen, Germany \\
$^2$Gesellschaft f\"ur Schwerionenforschung mbH, 
  Planckstr. 1  D-64291 Darmstadt, Germany \\
  $^3$ Dept. F\'isica Teor\'ica I, Universidad Complutense, 28040 Madrid, Spain }
\maketitle
\begin{abstract}
  We summarize our results for hadronic contributions to the anomalous magnetic 
  moment of the muon ($a_\mu$), the one from hadronic vacuum-polarisation (HVP) and the
  light-by-light scattering contribution (LBL), obtained from the Dyson-Schwinger equations (DSE's)
  of QCD. In the case of HVP we find good agreement with model independent determinations 
  from dispersion relations for $a_\mu^\mathrm{HVP}$ as well as for the Adler function 
  with deviations well below the ten percent level. From this we conclude that the DSE 
  approach should be capable of describing $a_\mu^\mathrm{LBL}$ with similar accuracy. 
  We also present results for LBL using a resonance expansion of the quark anti-quark 
  T-matrix. Our preliminary value is $a_\mu^\mathrm{LBL}=(217 \pm 91) \times 10^{-11}$.
\end{abstract}
\section{Introduction}
In the search for new physics beyond the standard model
the anomalous magnetic moment of the muon ($a_\mu$) is one of the most
interesting observables. Compared to the corresponding electron anomaly
($a_e$) it is more sensitive to contributions from high lying scales.
These include the weak interactions, QCD and potential new
physics \cite{Stockinger:2006zn}. Especially the contributions from soft QCD
desire highest attention because, due to their non-perturbative nature
and the resulting technical complications, they dominate the theoretical
standard model (SM) prediction. 

The efforts of the E821 experiment at Brookhaven National Lab
\cite{Bennett:2006fi,Roberts:2010cj} 
as well as theoretical efforts of more than a decade 
\cite{Jegerlehner:2009ry} culminated
in a determination of $a_\mu$ down to a level where significant
deviations have been found
      \begin{align}
		\label{eqn:amuexperiment}
            \mbox{Experiment:} \,\,\,\,
			&116\,592\,089(63)\times 10^{-11} \;\; , 
		\\
		\label{eqn:amutheoretical}
            \mbox{\phantom{wwu}} \mbox{Theory:} \,\,\,\,
			&116\,591\,828(49)\times 10^{-11} \;\; ,
      \end{align}
where the theoretical number is taken from Ref.~\cite{Hagiwara:2011af}.
Comparing theory and experiment the deviation amounts to $a_\mu^\mathrm{exp}-
a_\mu^\mathrm{theo}=261(80)$ which corresponds to a $3.2 \sigma$ effect.
In order to confirm this result the uncertainties have to be reduced further.

There are two hadronic contributions that dominate the SM uncertainty.
\begin{figure}[t!]
		\centering	\subfigure[][]{\label{fig:hadroniclo}\includegraphics[width=0.20\columnwidth]{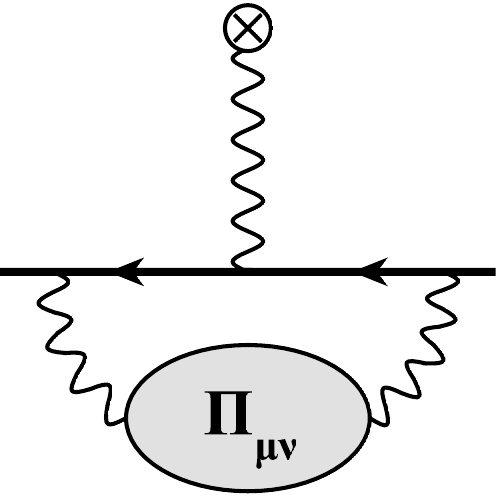}}
	\hspace{0.08\columnwidth}
	\subfigure[][]{\label{fig:hadroniclbl}\includegraphics[width=0.20\columnwidth]{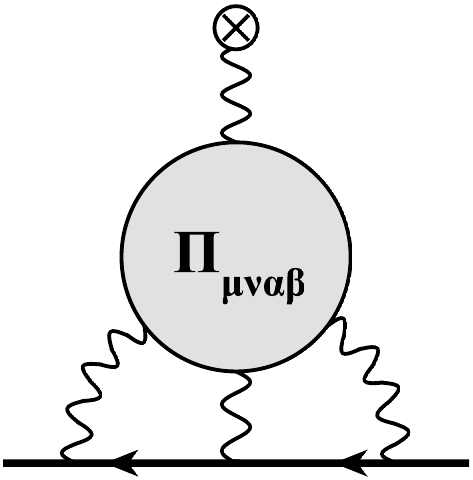}}
      \caption{The two classifications of corrections to the photon-muon
	vertex function: (a) hadronic vacuum polarization contribution to $a_\mu$. The vertex is 
               dressed by the vacuum polarization tensor $\Pi_{\mu\nu}$;
		   (b) the hadronic light-by-light scattering contribution to
		   $a_\mu$.}
\end{figure}
There is the hadronic vacuum polarisation contribution (HVP) which gives rise
to the leading hadronic contribution as well as the leading
SM uncertainty contribution \cite{Hagiwara:2011af} 
\begin{align}
  a_\mu^\mathrm{HVP,DR}=6\,949.1(42.7)\times 10^{-11}.
  \label{eqn:HVPLODispRel}
\end{align}
The relevant diagram, involving the hadronic one-particle irreducible (1PI) photon
self-energy $\Pi_{\mu\nu}$ is shown in figure \ref{fig:hadroniclo}.
The next-to leading uncertainty contribution comes from the light-by-light (LBL)
scattering contribution that is shown in fig. \ref{fig:hadroniclbl}. Estimates
from the viewpoint of effective field theory (EFT) from different approaches
were recently combined into a single number~\cite{Prades:2009tw}
\begin{align}
  a_\mu^\mathrm{LBL}=105(26)\times 10^{-11}.
  \label{eqn:PdRVlblresult}
\end{align}
The uncertainty given here is rather small compared to most estimates.
In fact our results indicate that this error may be far too optimistic.

Our strategy to determine these quantities is the following. 
We work with the  Dyson-Schwinger and
Bethe-Salpeter equations (DSE/BSE) of QCD \cite{Alkofer:2000wg,Fischer:2006ub}. 
With these we calculate the HVP contribution to $a_\mu$ where we 
use a parameter set, among others, that is completely fixed by meson
phenomenology. The HVP
contribution can be compared to essentially model independent result
from dispersion relations \cite{Eidelman:1998vc} such that the 
calculation serves as a non-trivial cross check of our methods.
Afterwards we approach the LBL contribution using exactly
the same truncation such that we have reasons to believe that we can ultimately
reach a similar precision as in the case of HVP.

This proceedings contribution is organized as follows. 
First of all we summarise the employed truncation in section
\ref{sec:Scheme}. The HVP
contribution will be discussed in section \ref{sec:HVP} and LBL
in sec. \ref{sec:LBL}. Afterwards we discuss our results for both
of these contributions in section
\ref{sec:Discussion} and conclude.

\section{Calculational scheme}
\label{sec:Scheme}
We work in rainbow-ladder truncation of QCD using the Maris-Tandy
model of the quark-gluon interaction \cite{Maris:1999nt}.
The central object in this approach is the quark DSE
\begin{align}
	  S^{-1}(p) = Z_2\,\,S_0^{-1}+Z_2^2\,\,\frac{4}{3}
	  \int_q \gamma_\mu S(p)\gamma_\nu\,\,T_{\mu\nu}(k)\, G(k^2),
	  \label{eqn:RLQuarkDSE}
\end{align}
where $S$ is the full quark propagator, $S_0$ the corresponding 
bare quantity and $Z_2$ is the quark wave-function renormalisation.
$T_{\mu\nu}(k)$ is the transverse projector and $G(k^2)$
is the effective gluon dressing. This function is modelled in the
present approach in a way such that chiral symmetry breaking occurs
while the axial-vector Ward-Takahashi identity (AXWTI), the $U(1)$ vector-WTI
of QED and resummed one-loop perturbation theory are respected \cite{Maris:1999nt}. 
Consistent with the quark DSE in (\ref{eqn:RLQuarkDSE}) is the meson BSE
\begin{align}
  [\Gamma]_{rs}(P,k)=  - Z_{2}^2\frac{4}{3}\int_q
  \,[S(q_+)\Gamma(P,q) S(q_-)]_{ut}K_{tu,rs}(k-q),
  \label{eqn:LadderMesonBSE}
\end{align}
where $P$ is the meson momentum, $k$ the relative
quark momentum and $q_\pm=q\pm P/2$.
The interaction kernel is defined as
\begin{align}
  K_{rs,tu}(k) &= G(k^2)
  T_{\mu\nu}(k) \big[ \gamma_\mu \big]_{rt}\big[ \gamma_\nu \big]_{us}.
  	  \label{RainbowLadderKernel}
\end{align}
The latter two equations are intimately related by chiral symmetry, to 
give a dynamical breaking  in accordance with
Goldstone's theorem \cite{Maris:1997hd}.
In addition we need the quark-photon vertex defined via the 
inhomogeneous BSE
\begin{align}
  [\Gamma_\mu]_{rs}(P,k)= Z_2\gamma_\mu - Z_{2}^2\frac{4}{3}\int_q
  \,[S(q_+)\Gamma_\mu(P,q) S(q_-)]_{ut}K_{tu,rs}(k-q),
  \label{eqn:LadderQEDVertexBSE}
\end{align}
which is the key to any calculation of electromagnetic
properties of hadrons.
The vertex features a vector-meson bound-state
for time-like momenta $P$ such that vector-meson dominance (VMD)
is dynamically included. This gives e.g. important
contributions to the pion charge radius which can be nicely
described in the present approach \cite{Maris:1999bh}. 

\section{Hadronic vacuum polarisation (HVP)}
\label{sec:HVP}
Here we present our results briefly, more details
can be found in Ref. \cite{Goecke:2011pe}. 
The central object for the HVP contribution is the hadronic
tensor
\begin{align}
  \Pi_{\mu\nu}(p) =-Z_2e^2 \int_q \mathrm{Tr} [S(q_+)\Gamma_\mu(p,q)S(q_-)\gamma_\nu ] \;\;,
  \label{eqn:HVPTensor}
\end{align}
which corresponds to the 1PI hadronic photon self-energy and involves the
non-perturbative quark propagator (\ref{eqn:RLQuarkDSE}) and the 
self-consistent quark-photon vertex (\ref{eqn:LadderQEDVertexBSE}).
The tensor $\Pi_{\mu\nu}$ is transverse due to its WTI

\begin{align}
  \Pi_{\mu\nu}(p) = \left(\delta_{\mu\nu}-\frac{p_\mu p_\nu}{p^2}\right)
  \,p^2\, \Pi(p^2) \;\; ,
  \label{eqn:HVPScalarDef}
\end{align}
which serves as a definition of the scalar function $\Pi(p^2)$. We use
the renormalisation condition $\Pi_\mathrm{R}(p^2) = \Pi(p^2)-\Pi(0)$
which corresponds to the usual physical QED on-shell scheme giving
rise to $e(0)=e_\mathrm{physical}$. Another quantity that is interesting
in the present context is the Adler function
\begin{align}
  D(q) = - q^2 \frac{d\,\Pi_\mathrm{R}(q^2)}{d\,q^2} \;\; .
  \label{eqn:AdlerFunctionDef}
\end{align}

\begin{figure}[t]
  \begin{center}
    \includegraphics[width=0.3\textwidth,angle=-90]{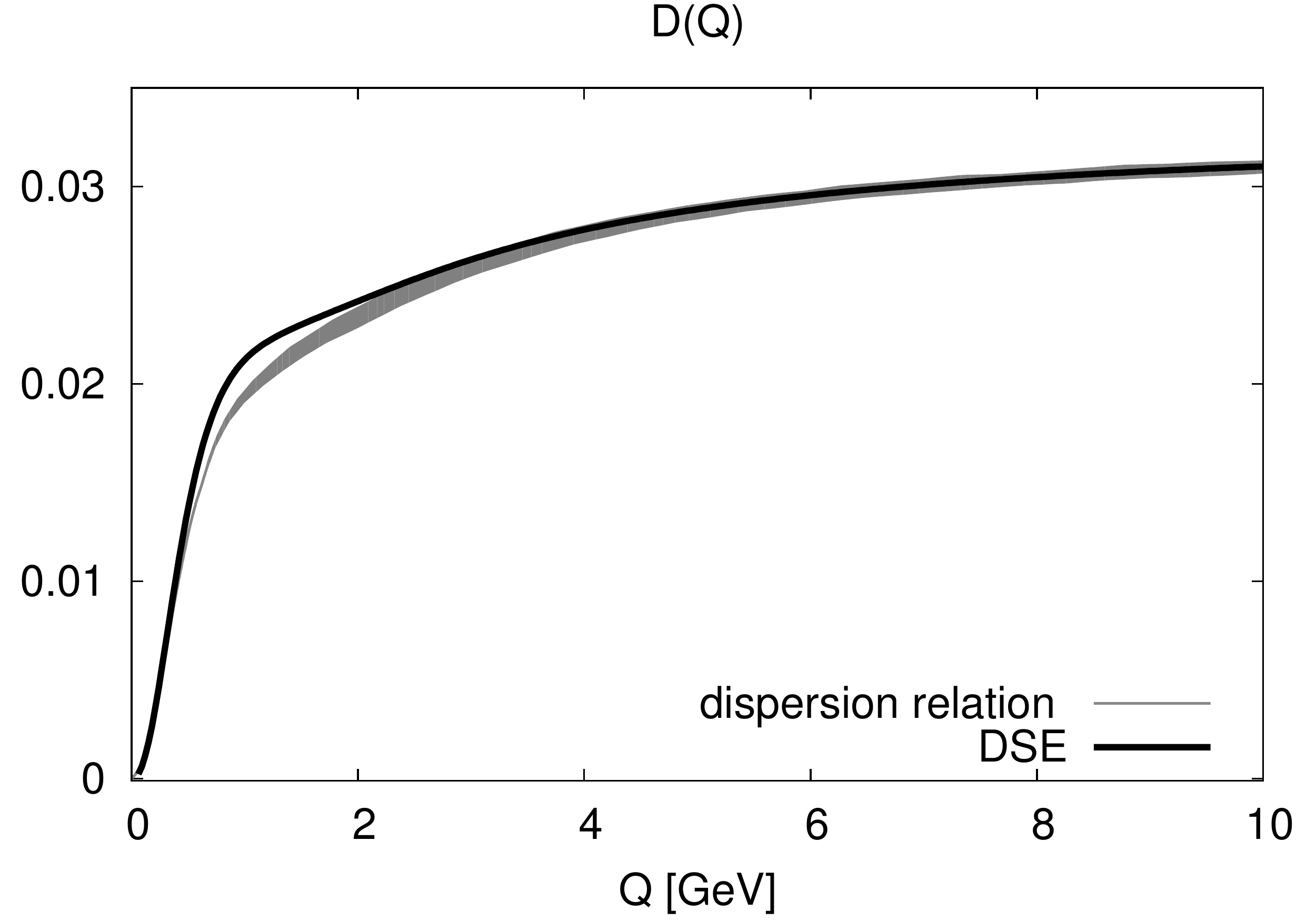}
  \end{center}
  \caption{The Adler function obtained from DSE's using our parameter set II compared to a model
  independent result from dispersion relations.}
  \label{fig:AdlerMainResult}
\end{figure}

The results for the Adler function from DSE's including five quark
flavours is shown in fig. \ref{fig:AdlerMainResult} and compared
to a result from dispersion relations \cite{Eidelman:1998vc}.
There is quite reasonable agreement at all momentum scales. 
Especially the deeply non-perturbative behaviour below $1$ GeV
is nicely reproduced. It is this regime, set by the muon mass, where the contribution
to $a_\mu$ saturates. 

The contribution to the muon $g-2$ can be obtained via \cite{deRafael:1993za} 
\begin{align}
  a_\mu^{\mathrm{HVP}}= \frac{\alpha}{\pi}\int_0^1\!\! dx\,\,(1-x) \left[-e^2\Pi_\mathrm{R}\left(\frac{x^2}{1-x}m_\mu^2  \right)  \right] \;\; ,
  \label{eqn:anomalyIntegral}
\end{align}
where $\alpha$ is the fine-structure constant. In particular we use
two different parameter sets. The standard parameter set where the $u$, $d$
and $s$ quark masses are fixed to the pseudo-scalar meson sector and another
one where instead the vector-meson sector is used. This is summarized in
table \ref{tab:param}.
\begin{table}[t]
  \centering
  \begin{tabular}{c|c|c|c|c|c|c}
    [MeV]	&	$m_{u,d}$	&	$m_{s}$ 	& 	$m_{\pi}$ & $m_{K}$ & $m_{\rho,\omega}$ & $m_{\phi}$	\\\hline\hline
	set I	&    $3.7$  	&	$85$ 		&	$138$	  &	$495$	&	$740$		 &	$1080$	\\\hline
	set II	&    $11$  		&	$72$ 		&	$240$	  &	$477$   &	$770$		 &	$1020$	\\\hline
  \end{tabular}
  \caption{Two choices for the light bare quark masses at $\mu^2=(19\,\mbox{GeV})^2$ and the resulting meson 
  masses (in MeV) in the pseudoscalar and vector meson sector. For the heavy quarks we always take $m_c = 827$ MeV and 
  $m_b = 3680$ MeV which lead to good results for charmonia and bottomonia in the pseudoscalar and vector channel.}
  \label{tab:param}
\end{table}
The $c$ and $b$ quark mass functions are fixed to the charmonium and
bottomonium
vector meson states in all cases \cite{Maris:2006ea}. With these two parameter sets
we obtain the following results
\begin{align}
  a^{{HVP,I}}_\mu &=7440\times 10^{-11}\;\; ,    \hspace*{1cm}
  a^{{HVP, II}}_\mu  = 6760 \times 10^{-11}\;\;. 
  \label{eqn:amuResults}
\end{align}
Comparing to the model independent result (\ref{eqn:HVPLODispRel})
we see that the standard set I deviates about 7 percent. We see the reason for this
in the $\rho$ mass which is about four percent too light, see tab. \ref{tab:param}.
The result with the physical $\rho$ mass (set II) is indeed closer to the dispersion
relation result. Taking the idea of changing the $\rho$ mass a step further,
we calculate the two flavour contribution $a_\mu^{\mathrm{HVP},N_f=2}$ as a function
of the vector meson mass. Both are functions of the quark masses $m_{u/d}$.
\begin{figure}[t]
  \begin{center}
    \includegraphics[width=0.3\textwidth,angle=-90]{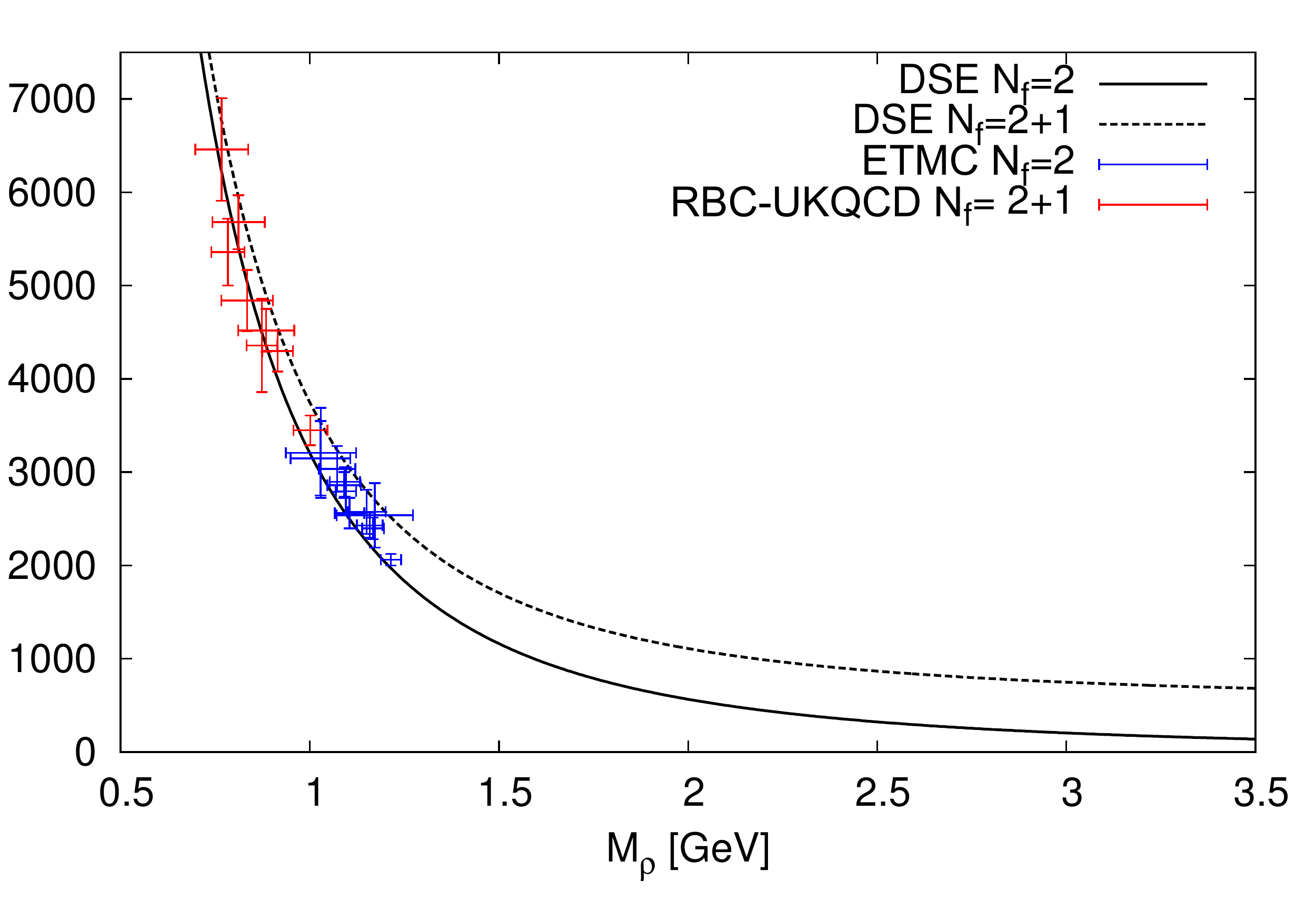}
  \end{center}
  \caption{The $N_f=2$ and $N_f=2+1$ flavour contribution to $a_\mu^\mathrm{HVP}$
  in units of $[10^{-11}]$ as
  a function of the vector meson mass. For the latter case the $s$ contribution is
  kept constant. The two DSE curves are compared to lattice results for the two- and
  three flavour case respectively. The blue data is $N_f=2$ data from ETMC 
  \cite{Feng:2011zk} 
  and the red data is $2+1$ from RBC-UKQCD \cite{Boyle:2011hu}.}
  \label{fig:a_vs_MV}
\end{figure}
In addition we calculate $a_\mu^{\mathrm{HVP},N_f=2+1}$ where the 
strange contribution is just an additive constant since $m_s$ remains fixed.
The results are shown in fig. \ref{fig:a_vs_MV} where we compare to
$N_f=2$ results from the ETMC collaboration \cite{Feng:2011zk} (blue data) and
$N_f=2+1$ data (red) from the RBC-UKQCD collaboration \cite{Boyle:2011hu}. 
The ordinate shows $a_\mu$ in units of $[10^{-11}]$. Our curves agree
with both data within error bars which we take as a hint that the DSE/BSE
approach in the present truncation captures the relevant degrees of freedom.
For HVP this seems to be more than anything else the vector meson as would
be expected from VMD estimates \cite{Gourdin:1969dm}. For a detailed discussion
see Ref. \cite{Goecke:2011pe}. 

\section{Hadronic light-by-light scattering (LBL)}
\label{sec:LBL}
In the present section we discuss the LBL contribution.
Within the framework of DSE's the hadronic four-point function,
that is the essential ingredient here, has a description that
is consistent with the one for HVP shown earlier (\ref{eqn:HVPTensor}).
We presented this representation in Refs. \cite{Fischer:2010iz,Goecke:2010if} 
where also more details can be found. There  we also elaborate on the resonance expansion
of the quark anti-quark T-matrix that is used for the results presented here. To this end
we arrive at an approximate description of the full four-point function
\begin{align}
  \parbox{3cm}{\includegraphics[width=2.2cm]{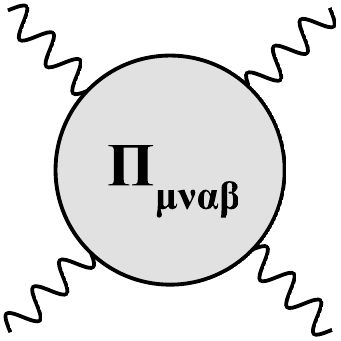}}
  &=\parbox{3cm}{\includegraphics[width=2.2cm]{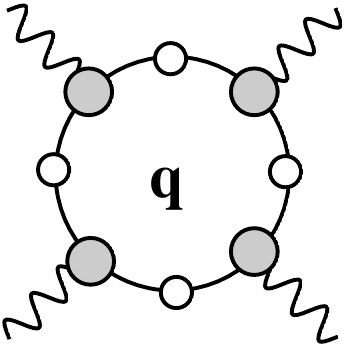}} 
  +\parbox{3cm}{\includegraphics[width=3cm]{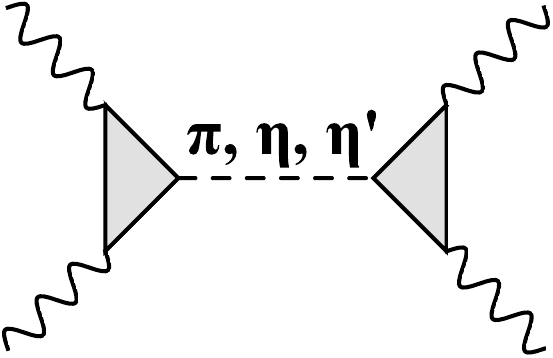}}, 
  \label{eqn:4ptResonanceExpansion}
\end{align}
that consists of the non-perturbatively dressed quark loop diagram (QL) as
well as a pseudo-scalar meson-exchange contribution that takes into account
the $\pi^0$, $\eta$ and $\eta^\prime$ mesons. This picture is very similar
to the one obtained in hadronic models and EFT   \cite{deRafael:1993za,Bijnens:1995cc, 
Hayakawa:1996ki,Hayakawa:1997rq,Knecht:2001qf,Melnikov:2003xd,Dorokhov:2008pw}.

In order to obtain the contribution to $a_\mu^\mathrm{LBL}$ from the four-point function
we define
	\begin{align}
  		ie\widetilde{\Gamma}_{\rho\mu} = \int_{q_1}\int_{q_2}&
			D_{\epsilon\nu}(q_1)D_{\delta\alpha}(q_2)D_{\gamma\beta}(q_3)
           (ie\gamma_\gamma) S(p_1) (ie\gamma_\delta) S(p_2) (ie\gamma_\epsilon)
   \left[  (ie)^4\widetilde{\Pi}_{(\rho)\mu\nu\alpha\beta}(q_1,q_2,q_3) \right],
	  \label{eqn:GammaRhoMu}
	\end{align}
from which the anomaly can be obtained via \cite{Aldins:1970id} 
	\begin{align}
	  a_\mu = \left.\frac{1}{48m_\mu}\mathrm{Tr}\left[(i\Sh{P}+m_\mu)[\gamma_\sigma,\gamma_\rho]
  		(i\Sh{P}+m_\mu) \widetilde{\Gamma}_{\sigma\rho}\right]\right|_{k\equiv0},
  		\label{eqn:DefinitionOfAnomaly}
	\end{align}
where $P$ is the muon momentum.
Here $S$ and $D_{\mu\nu}$ are perturbative muon and photon propagators and
the definition $\widetilde{\Pi}_{(\rho)\mu\nu\alpha\beta}=
\partial_{\rho}\Pi_{\mu\nu\alpha\beta}$ has been used with the hadronic four-point 
function $\Pi_{\mu\nu\alpha\beta}$. 

For the pseudo-scalar (PS) meson pole contribution we need the $\mathrm{PS}\gamma\gamma$
form factor 
	\begin{align}
            \Lambda_{\mu\nu}^{\mathrm{PS}\gamma\gamma}(k_1,k_2) &=\,
	    2e^2N_c\!\!\int_k \mathrm{Tr}\big[ i\hat{\mathcal{Q}}_e\Gamma_\nu(k_2,p_{12})S_F(p_2)	    \hat{\Gamma}^{\mathrm{PS}}(p_{23},P)S_F(p_3)i\hat{\mathcal{Q}}_e\Gamma_\mu(k_1,p_{31})S_F(p_3)\big]\;,
		\label{eqn:PseudoScalarFormFactor}
	\end{align}
that is defined as a non-perturbative quark triangle that involves the quark
(\ref{eqn:RLQuarkDSE}), the quark-photon vertex (\ref{eqn:LadderQEDVertexBSE})
and the meson amplitude (\ref{eqn:LadderMesonBSE}) called $\hat{\Gamma}^\mathrm{PS}$
here. The meson momentum is $P$, $k_{1/2}$ are the photon momenta, 
$p_i$ the momenta of the quarks
and $p_{ij}=\left(p_i+p_j\right)/2$. 
From the form factor together
with a bare meson propagator we obtain the resonant part that is shown in Eq.
(\ref{eqn:4ptResonanceExpansion}). Details concerning the flavour content of
the meson as well as the necessary meson off-shell prescription can be found
in \cite{Goecke:2010if}. Once the form factor is known
the contribution to $g-2$ can be obtained along the lines explained in \cite{Knecht:2001qf}. 
Our result for the PS meson exchange contribution
($\pi^0$, $\eta$, $\eta^\prime$) 
is
	\begin{align}
		a_\mu^{\textrm{LBL;PS}}=(80.7 \pm 12.0)\times 10^{-11}, \label{res:PS}
	\end{align}
where the error is dominantly an estimate of the systematic model uncertainty.

For the QL contribution we take the full quark propagator (\ref{eqn:RLQuarkDSE})
together with the Ball-Chiu (BC) vertex construction \cite{Ball:1980ay} 
\begin{align}
  \Gamma_\mu^{\mathrm{BC}}(P,k)&=
   \gamma_\mu \Sigma_A + 2\sh{k} k_\mu \Delta_A
   +i k_\mu \Delta_B,
  \label{eqn:BCVertex}
\end{align}
where the symbols 
\begin{xalignat}{2}
  \Sigma_F&=\frac{F(k_+^2)+F(k_-^2)}{2}
  &\Delta_F&=\frac{F(k_+^2)-F(k_-^2)}{k_+^2-k_-^2},
\end{xalignat}
have been used. This substructure of the fully self-consistent vertex
(\ref{eqn:LadderQEDVertexBSE}) is dictated by the vector WTI. Defining the four-point 
function from the quark-loop as in Eq. (\ref{eqn:4ptResonanceExpansion}),
taking the derivative and using Eqs. (\ref{eqn:GammaRhoMu}) and
(\ref{eqn:DefinitionOfAnomaly}) we obtain
	\begin{equation}
		\begin{array}{lcc}
		a_\mu^{\textrm{LBL;quarkloop (bare vertex)}} &=&
		(\phantom{0}61 \pm 2) \times 10^{-11}\\
		a_\mu^{\textrm{LBL;quarkloop (1BC)}}         &=& 
		(107 \pm 2)		    \times 10^{-11}\\
		a_\mu^{\textrm{LBL;quarkloop (BC)}}         &=& 
		(176 \pm 4)		    \times 10^{-11},\\
		\end{array}\label{res:QL} \\
		\end{equation}
where the first result uses a bare quark-photon vertex and $1BC$ only has
the $\gamma_\mu$ part of the dressed vertex, see (\ref{eqn:BCVertex}).
In this calculation we included the flavours $u$, $d$, $s$ and $c$. 
It can clearly be seen that the vertex dressing causes quite some enhancement
especially when all three structures of the BC vertex are used (third case).
The error given here is numerical. Clearly, this result is preliminary,
since the transverse structure of the vertex including important contributions 
from vector mesons is missing. We have estimated these contributions in 
\cite{Goecke:2010if}, however a full calculation is absolutely mandatory and 
well under way.

\section{Discussion}
\label{sec:Discussion}

We presented results for the HVP as well as for the LBL contribution to the muon
$g-2$ obtained within the framework of DSE's. We saw that in the case of HVP
our results for $a_\mu^\mathrm{HVP}$ (\ref{eqn:amuResults}) reproduce model
independent dispersion relations on the less-than-ten-percent level. We see no
principal reason why this should not be the case also for a full LBL calculation.
Indeed, our result for the pseudo-scalar meson-exchange contribution to
LBL (\ref{res:PS}) is in the ballpark of the results obtained
within other approaches \cite{deRafael:1993za,Bijnens:1995cc, 
Hayakawa:1996ki,Hayakawa:1997rq,Knecht:2001qf,Melnikov:2003xd,Dorokhov:2008pw}
(see \cite{Jegerlehner:2009ry} for an overview). For the quark loop
contribution to LBL we take our BC result from (\ref{res:QL}) and guesstimate
the missing contributions from the above mentioned transverse components of the 
quark-photon vertex, see \cite{Goecke:2010if} for details. Thus we arrive at the
value $a_\mu^{\textrm{LBL;quarkloop (BC+transverse)}} = (136 \pm 79)\times 10^{-11}$,
where the large error takes into account the uncertainties due to the estimate.
Puting all contributions together we obtain 
	\begin{equation}
		\begin{array}{lcc}
		a_\mu^{\textrm{LBL}} &=& (217 \pm 91)\times 10^{-11}\;, \\
		\end{array}
	\end{equation}
As mentioned already our result for LBL hints towards a larger contribution 
and thus to a smaller deviation between theory and experiment as compared to Eqs.
(\ref{eqn:amuexperiment}, \ref{eqn:amutheoretical}). Besides implementing the full
quark-photon vertex inside the quark-loop, another important next step is to
overcome the resonance approximation of LBL.
In general, we believe to have shown that a full calculation 
from the DSE approach can be expected to mark a clear step
forward for the case of the hadronic light-by-light scattering
contribution to the muon $g-2$.

\section{Acknowledgements}   
This work was supported by the DFG under grant No.~Fi 970/8-1, by the 
Helmholtz-University Young Investigator Grant No.~VH-NG-332 and by the 
Helmholtz International Center for FAIR within the LOEWE program of the 
State of Hesse. 
RW would also like to acknowledge support by the Austrian
Science Fund FWF under Project No. P20592-N16, and by 
Ministerio de Educaci\'on (Spain): Programa Nacional de Movilidad de 
Recursos Humanos del, Plan Nacional de I-D+i 2008-2011.

\end{document}